\documentclass[aps,prb,twocolumn,groupedaddress]{revtex4}
\usepackage{graphicx}

\usepackage{amsmath}

\newcommand{\leco}{La$_{1-x}$Eu$_x$CoO$_3$}
\newcommand{\eco}{EuCoO$_3$}
\newcommand{\lco}{LaCoO$_3$}
\newcommand{\lsco}{La$_{1-x}$Sr$_x$CoO$_3$}

\newcommand{\lmco}{La$_{1-x}M_x$CoO$_3$}
\newcommand{\lcco}{La$_{1-x}$Ca$_x$CoO$_3$}

\newcommand{\ch}{susceptibility}

\newcommand{\tad}{thermal expansion}

\newcommand{\sst}{spin-state transition}
\newcommand{\mi}{metal-insulator transition}
\newcommand{\tm}{$T_{\rm MI}$}
\newcommand{\CF}{$\Delta_{\rm CF}$}
\newcommand{\JH}{$J_{\rm H}$}

\newcommand{\tg}{$t_{2g}$}
\newcommand{\eg}{$e_{g}$}
\newcommand{\LS}{$t_{2g}^{\,6}e_{g}^0$}
\newcommand{\IS}{$t_{2g}^{\,5}e_{g}^1$}
\newcommand{\HS}{$t_{2g}^{\,4}e_{g}^2$}

\begin{document}
%\preprint{}

\title{Spin-State Transition and Metal-Insulator Transition in La$_{1-x}$Eu$_x$CoO$_3$}

\author{J.\,Baier, S.\,Jodlauk, M.\,Kriener, A.\,Reichl,
C.\,Zobel, H.\,Kierspel, A.\,Freimuth, and T.\,Lorenz}

\affiliation{ II.\,Physikalisches Institut, Universit\"{a}t zu K\"{o}ln,
Z\"{u}lpicher Str. 77, 50937 K\"{o}ln, Germany}

\date{\today}

\begin{abstract}

We present a study of the structure, the electric resistivity, the
magnetic susceptibility, and the thermal expansion of
La$_{1-x}$Eu$_x$CoO$_3$. LaCoO$_3$ shows a temperature-induced
spin-state transition around 100\,K and a metal-insulator
transition around 500\,K. Partial substitution of La$^{3+}$ by
the smaller Eu$^{3+}$ causes chemical pressure and leads to a
drastic increase of the spin gap from about 190\,K in \lco\ to
about 2000\,K in \eco , so that the spin-state transition is
shifted to much higher temperatures. A combined analysis of \tad\
and \ch\ gives evidence that the \sst\ has to be attributed to a
population of an intermediate-spin state without orbital
degeneracy for $x<0.5$ and with orbital degeneracy  for larger
$x$. In contrast to the \sst ,  the metal-insulator transition is
shifted only moderately to higher temperatures with increasing Eu
content, showing that the metal-insulator transition occurs
independently from the spin-state distribution of the Co$^{3+}$
ions. Around the \mi\ the magnetic \ch\ shows a similar increase
for all $x$ and approaches a doping-independent value around
1000\,K indicating that well above the \mi\ the same spin state
is approached for all $x$.

\end{abstract}

%\pacs{71.30.+h, 72.80.Ga, 71.27.+a}

\maketitle

\section{Introduction}

\lco\ has unusual physical properties, which have attracted
continuing interest for decades.\cite{jonker53a} Nevertheless, the
magnetic, electronic, and structural properties of this compound
remain the subject of controversial discussion. The magnetic
susceptibility $\chi$ shows a maximum around 100\,K that is usually
ascribed to a spin-state transition from a nonmagnetic insulating
state at low temperatures to a paramagnetic insulating state at
higher
temperatures.\cite{raccah67a,bhide72a,asai94a,itoh94a,korotin96a,potze95a,
saitoh97b,asai98a,yamaguchi97a,kobayashi00b} In addition, around
500\,K a \mi\ occurs,\cite{senaris95a,tokura98a} which is
accompanied by an increase of the magnetic moment leading to a broad
plateau in the magnetic susceptibility.\cite{yamaguchi96a,zobel02a}
The Co$^{3+}$ ions may occur in three different spin states, the
low-spin (LS)(\LS , $S=0$), the intermediate-spin (IS)(\IS , $S=1$),
and the high-spin (HS) state (\HS , $S=2$). The energies of these
spin states depend on the balance between the crystal field
splitting (\CF ) and the Hund's rule coupling (\JH ). A multiplet
calculation for a $3d^6$ configuration in a cubic crystal field
yields that the ground state is either the LS (large \CF ) or the HS
state (large \JH ), but never the IS state.\cite{sugano70}
Therefore, in earlier publications the \sst\ was often attributed to
a thermal population of the HS state from the LS ground
state.\cite{raccah67a,asai94a,itoh94a,senaris95a,yamaguchi96a}
However, a population of the HS state should result in a
susceptibility that is much larger than observed
experimentally.\cite{zobel02a} More recent
investigations~\cite{potze95a,saitoh97b,asai98a,yamaguchi97a,
kobayashi00b,zobel02a,radaelli02a} favor a LS/IS scenario, which
yields a quantitative description of $\chi (T)$ and which also
explains an experimentally observed scaling behavior between $\chi
(T)$ and the \tad\ $\alpha(T)$.\cite{zobel02a} From a theoretical
point of view the LS/IS scenario is supported by the results of
LDA+U calculations (where LDA stands for the local density
approximation), which yield that the IS state may be stabilized by a
hybridization between Co \eg\ and O $2p$
levels.\cite{korotin96a,nekrasov03a} Nevertheless, the question of
which spin state is populated is not yet clarified unambiguously.
Based on high-field electron spin resonance (ESR) data the
population of a spin-orbit coupled HS state with a total angular
momentum $\tilde{j}=S-\tilde{l}=1$ has been suggested
recently.\cite{noguchi02a} Here, $\tilde{l}=1$ denotes the effective
orbital momentum arising from the partial occupation of the \tg\
orbital. However, this ESR result is in conflict with the \ch\ data,
because the ESR data yield a $g$ factor of about 3.5 which leads to
a magnetic \ch\ that is about three times larger than that observed
experimentally.

Within the originally proposed LS/HS scenario the reduced absolute
value of the \ch\ has been traced back to an ordering of LS and HS
Co$^{3+}$ ions. Moreover, this ordering was used to explain that
despite the partial occupation of \eg\ states above the \sst\ around
100\,K \lco\ remains insulating up to about 500\,K and the \mi\ was
attributed to a melting of this order. However, no experimental
evidence for such a LS/HS superstructure has been reported so far.
Within the LS/IS scenario the occurrence of orbital order which
remains stable up to 500\,K has been proposed as a cause why the
\mi\ and the \sst\ do not coincide.\cite{korotin96a} Experimental
evidence for orbital order stems from changes of phonon modes
measured by optical spectroscopy\,\cite{yamaguchi97a} and from a
pair density function analysis of pulsed neutron
data\,\cite{louca99a}. More recently, a high-resolution single
crystal x-ray study of \lco\ has revealed a small monoclinic
distortion that is related to the Jahn-Teller (JT) effect of the
thermally excited Co$^{3+}$ ions in the IS state.\cite{maris03a}
Indirect evidence for this scenario is also obtained from the
scaling between $\chi (T)$ and $\alpha(T)$ around the \sst
,\cite{zobel02a} which requires that the orbital degeneracy of the
IS state is lifted as it is the case for JT-distorted CoO$_6$
octahedra. Moreover, a suppression of the JT distortion in the
metallic phase should restore the orbital degeneracy and thus could
partially explain the enhanced magnetic susceptibility above the \mi
.\cite{zobel02a} Alternative explanations of the enhanced \ch\ in
the metallic phase rely on scenarios which include all three
different spin states, assuming that the IS state is populated
around the \sst\ and the HS state becomes populated at the \mi
.\cite{saitoh97b,asai98a,radaelli02a}

In order to get more insight to the \sst\ of \lco\ and its relation
to the \mi\ we present an experimental study of the structural
changes, the magnetic susceptibility, the electrical resistivity,
and the thermal expansion of \leco\ with $0\le x\le 1$. The basic
idea is that the partial substitution of La$^{3+}$ by the smaller
Eu$^{3+}$ systematically increases the chemical pressure which
should enhance the crystal field splitting and therefore stabilize
the low-spin state. We have chosen the substitution by Eu$^{3+}$
because it has a vanishing total angular momentum $J$ of the 4f
shell. Using other rare earth ions with finite $J$ would strongly
hamper the analysis of the Co$^{3+}$ magnetism. Our main findings
are as follows: (i) With increasing Eu content the spin-state
transition is drastically shifted towards higher temperatures. (ii)
We observe a scaling behavior between the magnetic \ch\ and the \tad
, which gives clear evidence that for low temperatures ($T<200$\,K)
and low Eu content ($x\le 0.25$) the \sst\ arises from a thermal
population of the IS state without orbital degeneracy. The lack of
orbital degeneracy can be interpreted as orbital order or as a
consequence of JT distortions of the CoO$_6$ octahedra with
Co$^{3+}$ in the JT-active IS state~\cite{OOvsJT}. In addition, we
find some indication that there is an orbital degeneracy of the IS
state for $x>0.5$. (iii) All samples are good insulators below
400\,K. The electrical resistivity behaves like $\rho \propto
\exp(\Delta_{\rm act}/T)$, with an activation energy $\Delta_{\rm
act}$ that strongly increases with Eu content. (iv) The \mi\ is
shifted only moderately to higher temperatures \tm\ with increasing
Eu content, much weaker than that of the \sst . This shows that \tm\
is only weakly influenced by the thermal population of the IS state.
(v) For all Eu concentrations the \mi\ is accompanied by an increase
of $\chi$ of about $5\cdot 10^{-4}$\,emu/mole and the \ch\
approaches a common value for $T\rightarrow 1000$\,K indicating a
common spin state. The nature of this high-temperature spin state
remains, however, unclear, because it cannot be deduced from an
analysis of the \ch\ data alone.

\section{Experiment}

We prepared crystals of \leco\ with $x=0$, 0.1, 0.15, 0.2, 0.25,
0.5, 0.75, and 1 by the floating zone method in a four-mirror image
furnace. The preparation parameters are essentially the same as
those described in Ref.\,\onlinecite{kriener04a} for the growth of
\lsco\ single crystals. The LaCoO$_3$ single crystal studied here is
the same as used in Refs.\,\onlinecite{zobel02a,kriener04a}. The
stoichiometry of the crystals has been checked by energy-dispersive
x-ray analysis. All samples are single phase as has been confirmed by
x-ray powder diffraction at room temperature. From Laue photographs
of both ends of the crystals we have found that \lco\ and \eco\ are
single crystals, whereas the samples with $0<x<1$ are
polycrystalline. The electrical resistivity has been measured by a
standard four-probe technique on bar-shaped samples of typical
dimensions of $2\times 3 \times 5$\,mm$^3$. Wires have been attached
by conductive silver epoxy to four gold stripes, which have been
evaporated on the sample surface. We have used a dc technique for
resistivity values above 1\,$\Omega$ and an ac technique in the
sub-$\Omega$ range. The susceptibility measurements have been
performed between 4 and 300\,K in a field of 0.05\,T using a
superconducting quantum interference device magnetometer
(\textsl{Cryogenic}, S600X) and in the temperature range from 200 to
1000\,K in a field of 1\,T with a home-built Faraday balance.
High-resolution measurements of the linear thermal expansion were
performed using a home-built capacitance dilatometer in the
temperature range from 4 to 180\,K.\cite{pott83a}

\section{Results}

The results of our room temperature powder x-ray diffraction
measurements of \leco\ are summarized in Table~\ref{tab:lattice}.
With increasing Eu content, the cell volume per formula unit shrinks
almost linearly by about 6\,\% from 56\,\AA$^3$ in \lco\ to
52.75\,\AA$^3$ in \eco . Between $x=0.2$ and $0.25$ the symmetry
changes from rhombohedral $(R\overline{3}c)$ to orthorhombic
$(Pnma)$, as it is typical for perovskites when the tolerance factor
$t=(\langle r_{\rm A}\rangle+r_{\rm O})/(\sqrt{2}(r_{\rm Co}+r_{\rm
O}))$ decreases. Here $r_{\rm Co}$ and $r_{\rm O}$ denote the ionic
radii of the Co$^{3+}$ and O$^{2-}$ ions, respectively, and $\langle
r_{\rm A}\rangle=(1-x)\,r_{\rm La}+x\,r_{\rm Eu}$ is the average
radius of the A site ions, i.\,e.\ the average radius of La$^{3+}$
and Eu$^{3+}$. Such a symmetry change from rhombohedral to
orthorhombic has also been observed in \lcco\ for $x\ge
0.2$.\cite{kriener04a,burley04a}

\begin{table}[b]%[H] add [H] placement to break table across pages
 \caption{Room temperature lattice constants and volume per
 formula unit of \leco . The symmetry is rhombohedral
 ($R\bar{3}c$) for $x\le 0.2$ and
 orthorhombic (Pnma) for $x\ge 0.25$.
 \label{tab:lattice}}
 \vskip1mm
 \parbox{7cm}{
 \begin{ruledtabular}
 \begin{tabular}{ccccccc}
 $x$ & $a_{\rm R}$\,(\AA) & $\alpha_{\rm R}$ & $a$\,(\AA)
     & $b$\,(\AA) & $c$\,(\AA) & $V$\,(\AA$^3$/f.u.)\\
 \hline
 %\vspace{3pt}
 0  &5.379 & $60.8^\circ$ & & &                & 56.02 \\
 0.1 &5.367 & $60.9^\circ$ & & &                & 55.76 \\
 0.15 &5.393 & $60.4^\circ$ & & &                & 55.99 \\
 0.2  &5.387 & $60.3^\circ$ & & &                & 55.60 \\
 0.25 &      &              &5.409 & 7.611& 5.370& 55.25 \\
 0.5 &      &              &5.353 & 7.565& 5.344& 54.10 \\
 0.75 &      &              &5.357 & 7.531& 5.314& 53.60 \\
 1  &      &              &5.370 & 7.477& 5.255& 52.75 \\
 \end{tabular}
 \end{ruledtabular}}
 \end{table}

\subsection{Resistivity}

\begin{figure}
\includegraphics[angle=0,width=8cm,clip]{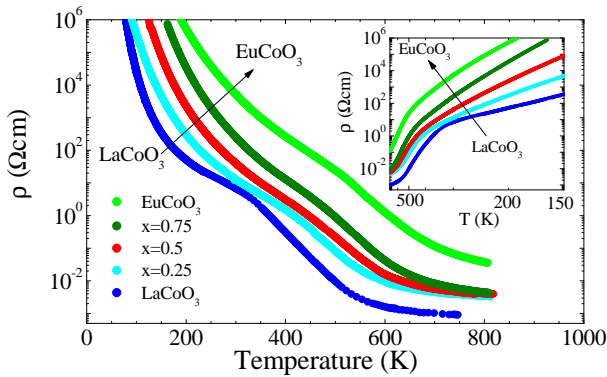}
\caption{Electrical resistivity of La$_{1-x}$Eu$_x$CoO$_3$ as a
function of temperature for different $x$. The inset shows the
same data on a reciprocal temperature scale.\label{fig:rho1}}
\end{figure}

In Fig.\,\ref{fig:rho1} we present the electrical resistivity $\rho$
of \leco . All samples studied are insulators. We find a systematic
increase of $\rho$ with increasing Eu content. Qualitatively, such
an increase may be expected from the decreasing tolerance factor.
When $t$ decreases, the deviation of the Co-O-Co bond angle from
180\,$^\circ$ increases and therefore the hopping probability of
charge carriers decreases. In accord with
Refs.\,\onlinecite{tokura98a,yamaguchi96a,english02a,mahendiran96a}
the resistivity of \lco\ shows an activated behavior ($\rho \propto
\exp(\Delta_{\rm act}/T)$) below 400\,K with an activation energy
$\Delta_{\rm act}\simeq 1200$\,K (see also
Ref.\,\onlinecite{kriener04a}). Above 400\,K a steep decrease of the
resistivity takes place, which is, according to optical conductivity
data~\cite{tokura98a}, attributed to a \mi\ around 500\,K. Note that
above this transition LaCoO$_3$ remains a rather poor metal with a
resistivity of about 1\,m$\Omega$cm. The Eu-doped samples show
similar $\rho(T)$ curves. As shown in the inset of
Fig.\,\ref{fig:rho1} the slopes of the $\log(\rho)$ versus $T^{-1}$,
i.\,e.\ $\Delta_{\rm act}$, curves strongly increase with increasing
Eu content. The onset of the steep decrease of $\rho(T)$ is shifted
to higher temperatures with increasing $x$, signaling that the \mi\
is shifted to higher temperatures.\cite{IIvsMI} However, the shift
of the transition temperature \tm\ is much weaker than that of
$\Delta_{\rm act}$. As will be shown later (Fig.\,\ref{fig:chirho1})
$\Delta_{\rm act}$ increases by about a factor of 3 and \tm\ only by
about 25\,\% from $x=0$ to~1. Such a different increase of
$\Delta_{\rm act}$ and \tm\ has also been found in resistivity
measurements on $R$CoO$_3$ with $R\,=\,La, \ldots,
Gd$.\cite{yamaguchi96b}

\subsection{Magnetic susceptibility}

\begin{figure}
\includegraphics[angle=0,width=8cm,clip]{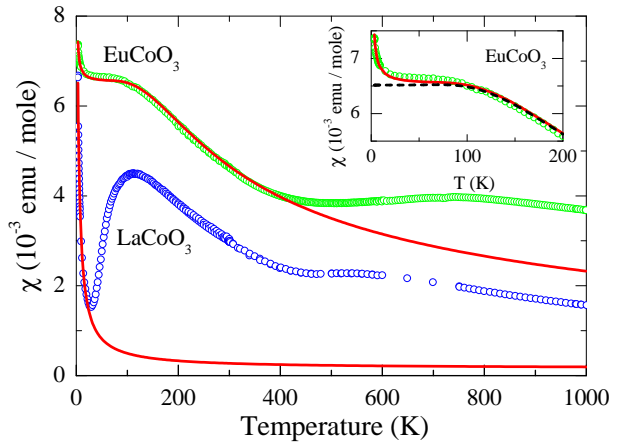}
\caption{Magnetic susceptibility of \eco\ and \lco\ as a function
of temperature (symbols) together with fits (solid lines) of the
respective background contributions (Eq.\,\ref{eq:background}).
The inset shows an expanded view of the low-temperature \ch\ of
\eco , which is almost entirely given by the Eu$^{3+}$ van Vleck
contribution shown by the dashed line. \label{fig:fitcdef}}
\end{figure}

In Fig.\,\ref{fig:fitcdef} we compare the magnetic \ch\ of \lco\ and
of \eco . In order to extract the Curie \ch\ of the Co$^{3+}$ ions
from the raw data we subtract a background contribution consisting
of (1) a Curie-Weiss contribution $\chi_{\rm
imp}=\frac{C}{T-\Theta}$ due to magnetic impurities and/or oxygen
nonstoichiometry, (2) a temperature-independent contribution
$\chi_0$ due to the diamagnetism of the core electrons and the van
Vleck susceptibility of the Co$^{3+}$ ions,\cite{remarkCovV} and (3)
the van Vleck susceptibility $\chi_{\rm vV}^{\rm Eu}$ of the
Eu$^{3+}$ ions weighted by the Eu content $x$. Note that it is
\textsl{a priori} not clear that in \leco\ all Eu ions are
three-valent. However, due to the $4f^7$ configuration of divalent
Eu$^{2+}$ with S\,=\,7/2 its presence would cause a strong increase
of the Curie-Weiss contribution $\chi_{\rm imp}$. Since this is not
at all the case in our samples, we conclude that the amount of
Eu$^{2+}$ is negligibly small. The temperature dependence of the van
Vleck susceptibility of Eu$^{3+}$ ions is well known and depends
solely on the energy gap $\Delta_{\rm Eu}$ between the $J=L-S=0$
ground state and the $J=1$ first excited state of the $4f^6$
multiplet of Eu$^{3+}$.\cite{vanvleck65} The gap $\Delta_{\rm Eu}$
is typically of the order of 500\,K. The entire background
susceptibility of \leco\ can thus be written as
\begin{equation}\label{eq:background}
  \chi_{\rm bg}(T)=\frac{C}{T-\Theta}+\chi_0+x\cdot\chi_{\rm vV}^{\rm Eu}(\Delta_{\rm Eu},T)\,.
\end{equation}
The parameters $C$, $\Theta$, $\chi_0$, and $\Delta_{\rm Eu}$ are
determined by nonlinear least-square fits of $\chi_{\rm bg}$ to the
low-temperature data, where the Co$^{3+}$ ions are in the
nonmagnetic LS state. As shown in Fig.\,\ref{fig:fitcdef},
$\chi_{\rm bg}$ describes the low-temperature $\chi(T)$ up to about
25\,K for \lco\ and about 400\,K for \eco .\cite{fitrange} This
implies that the Co$^{3+}$ ions remain essentially in the LS state
up to these temperatures, respectively.

\begin{table}[b]%[H] add [H] placement to break table across pages
 \caption{Fit parameters of the background susceptibility
 $\chi_{\rm bg}(T)$ (see Eq.\,\ref{eq:background}),\cite{Eugap} which describes the
 low-temperature data of the measured susceptibility. \label{tab:FitcDef}}
 \vskip1mm
 \parbox{7cm}{
 \begin{ruledtabular}
 \begin{tabular}{lcccc}
 Eu content & $\chi_0$ ($10^{-4} \frac{\rm emu}{\rm mole}$)
 & C ($\frac{\rm emuK}{\rm mole}$) & $\Theta$ (K) & $\Delta_{\rm Eu}$(K) \vspace{3pt}\\
 \hline
 \vspace{3pt}
 LaCoO$_3$&1.6 & 0.034 & -2.2 & --  \\
 $x=0.1$& 0 & 0.046 & -3.4 &460 \\
 $x=0.15$& 1.0 & 0.022 & -2.8 &460 \\
 $x=0.2$& 1.0 & 0.016 & -2.1 &460 \\
 $x=0.25$& 1.0 & 0.012 &-2.8& 460 \\
 $x=0.5$& 0.3 & 0.01 & 0& 462 \\
 $x=0.75$& 0 & 0.012 & 0&473 \\
 EuCoO$_3$& 0 & 0.003 & 0&457 \\
 \end{tabular}
 \end{ruledtabular}}
 \end{table}

\begin{figure}
\includegraphics[angle=0,width=8cm,clip]{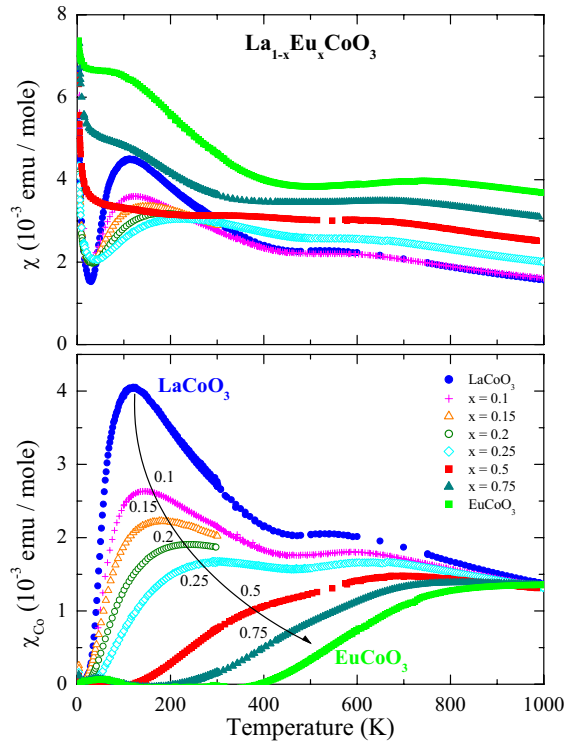}
\caption{The upper panel presents the magnetic susceptibility of
\leco\ as a function of temperature. With increasing Eu content
the susceptibility at low temperatures is more and more dominated
by the van Vleck contribution of the Eu$^{3+}$ ions. The lower
panel shows the Curie susceptibility of the Co$^{3+}$ ions after
subtracting a background \ch\ according to
Eq.\,\ref{eq:background} (see text). \label{fig:chichico}}
\end{figure}

In the upper panel of Fig.\,\ref{fig:chichico} we show the
susceptibility of La$_{1-x}$Eu$_{x}$CoO$_3$ for $0\le x\le 1$. When
the Eu content increases the \ch\ at low temperatures becomes more
and more dominated by the Eu$^{3+}$ van Vleck contribution. We have
fitted the respective low-temperature $\chi(T)$ data by
Eq.\,(\ref{eq:background}) for all samples and summarized the fit
parameters in Table~\ref{tab:FitcDef}. We find $\Delta_{\rm Eu}
\simeq 460$\,K (Ref.\,\onlinecite{Eugap}) and for the impurity
contribution we have obtained $C\simeq 0.02$\,emuK/mole and
$\Theta\simeq -3$\,K corresponding to impurity concentrations of
less than 1\,\%~(see Ref.\,\onlinecite{zobel02a}) with weak
antiferromagnetic interaction. For the sum of core diamagnetism and
Co$^{3+}$ van Vleck paramagnetism we find $0\le \chi_0 \le 1.6\cdot
10^{-4}$\,emu/mole. In our previous analysis of $\chi(T)$ of \lco\
we have used slightly different parameters, namely $\chi_0=6.5\cdot
10^{-4}$\,emu/mole, $C=0.02$\,emuK/mole and
$\Theta=0$.\cite{zobel02a} The main difference concerns the smaller
value of $\chi_0$, which we use in the present analysis for the
following reason: The low-temperature $\chi(T)$ data of all \leco\
samples with $x > 0$ can be well reproduced only for $\chi_0 \le
10^{-4}$\,emu/mole. Since it appeared very unlikely that $\chi_0$ is
significantly larger for $x=0$ than for all samples with $x>0$, we
have measured the magnetization of \lco\ at 1.8\,K up to 14\,T. This
field is sufficient to saturate the impurity contribution of the
magnetization and the remaining slope of $M(H)$ in the high-field
region, e.\,g.\ for $H\ge 11$~T where  $\mu_{\rm B}H /k_{\rm B} T
\ge 4$, is then given by $\text{d}M/\text{d}H \simeq \chi_0 $. For
\lco , this method yields $\chi_0 = 1.6\cdot 10^{-4}$\,emu/mole,
which is a much more reliable measure of $\chi_0$ than the fit of the
low-temperature \ch\ data by Eq.\,(\ref{eq:background}) because of
the rather restricted low-temperature range for $x=0$ ($T\lesssim
25$\,K).

In the lower panel of Fig.\,\ref{fig:chichico} we show the Curie
susceptibility of the Co$^{3+}$ ions after subtracting $\chi_{\rm
bg}$ from the raw data. The \ch\ of  \lco\ shows a low-temperature
maximum due to the \sst\ and a high-temperature shoulder around
the \mi . With increasing Eu content the low-temperature maximum
is continuously suppressed and strongly shifted to higher
temperatures, whereas the high-temperature shoulder moves only
moderately to higher temperatures corresponding to the weak
increase of \tm\ with $x$ inferred from $\rho (T)$. For $x<0.75$
two separate anomalies are clearly seen in the $\chi_{\rm Co}(T)$
curves, while in \eco\ these anomalies have finally merged into
one broad increase of $\chi(T)$. It is remarkable that after the
subtraction of $\chi_{\rm bg}$ the \ch\ data of all samples
approach a value of about $1.4\cdot 10^{-3}$\,emu/mole for
$T\rightarrow 1000$\,K.

Without further quantitative analysis the data in the lower panel of
Fig.\,\ref{fig:chichico} clearly show that the \sst\ is
systematically shifted towards higher temperatures with increasing
Eu content. Similar to physical pressure in
LaCoO$_3$,\cite{asai97a,vogt03a} the chemical pressure imposed by Eu
stabilizes the LS state of the Co$^{3+}$ ions due to the enhanced
crystal-field splitting. This conclusion agrees with the results of
Co nuclear magnetic resonance (NMR) studies on $R$CoO$_3$ with $
R\,=\,La, \ldots, Eu$.\cite{itoh99a,itoh00b} A quantitative analysis
of $\chi_{\rm Co}(T,x)$ and its relation to the \mi\ will be given
below.

\subsection{Thermal expansion}

In Fig.\,\ref{fig:thermex1} we present high-resolution measurements
of the linear thermal expansion $\alpha$ of \leco . As shown in
Ref.\,\onlinecite{zobel02a}, $\alpha$ is  a very sensitive probe of
the \sst . In \lco\ $\alpha$ steeply increases above about 25\,K and
reaches a maximum around 50\,K. This unusual behavior can be
attributed to the \sst , which leads to a thermal population of the
\eg\ orbitals. Since the \eg\ orbitals are oriented towards the
surrounding negative O$^{2-}$ ions, this causes an anomalous,
additional contribution $\Delta \alpha$ to the lattice expansion. As
reported in Ref.\,\onlinecite{zobel02a} the \tad\ and the magnetic
\ch\ fulfill the scaling relation
\begin{equation}
\label{scala}
 C \cdot \Delta \alpha(T) = \frac{\partial (\chi_{\rm Co}\cdot T)}{\partial T}
\end{equation}
with a scaling factor $C$ depending on the spin state of the excited
state. For LaCoO$_3$, we find C $simeq$ 195 emu K/mole, which agrees
very well with the value expected for a population of the IS state
without orbital degeneracy (190 emu K/mole) and strongly deviates
from the values of other scenarios such as a population of the IS
state with orbital degeneracy (290 emu K/mole) or a thermal
population of the HS state with (1000 emu K/mole) and without
orbital degeneracy (690 emu K/mole),
respectively.\cite{zobel02a,remark_C}

\begin{figure}[t]
\includegraphics[angle=0,width=8cm,clip]{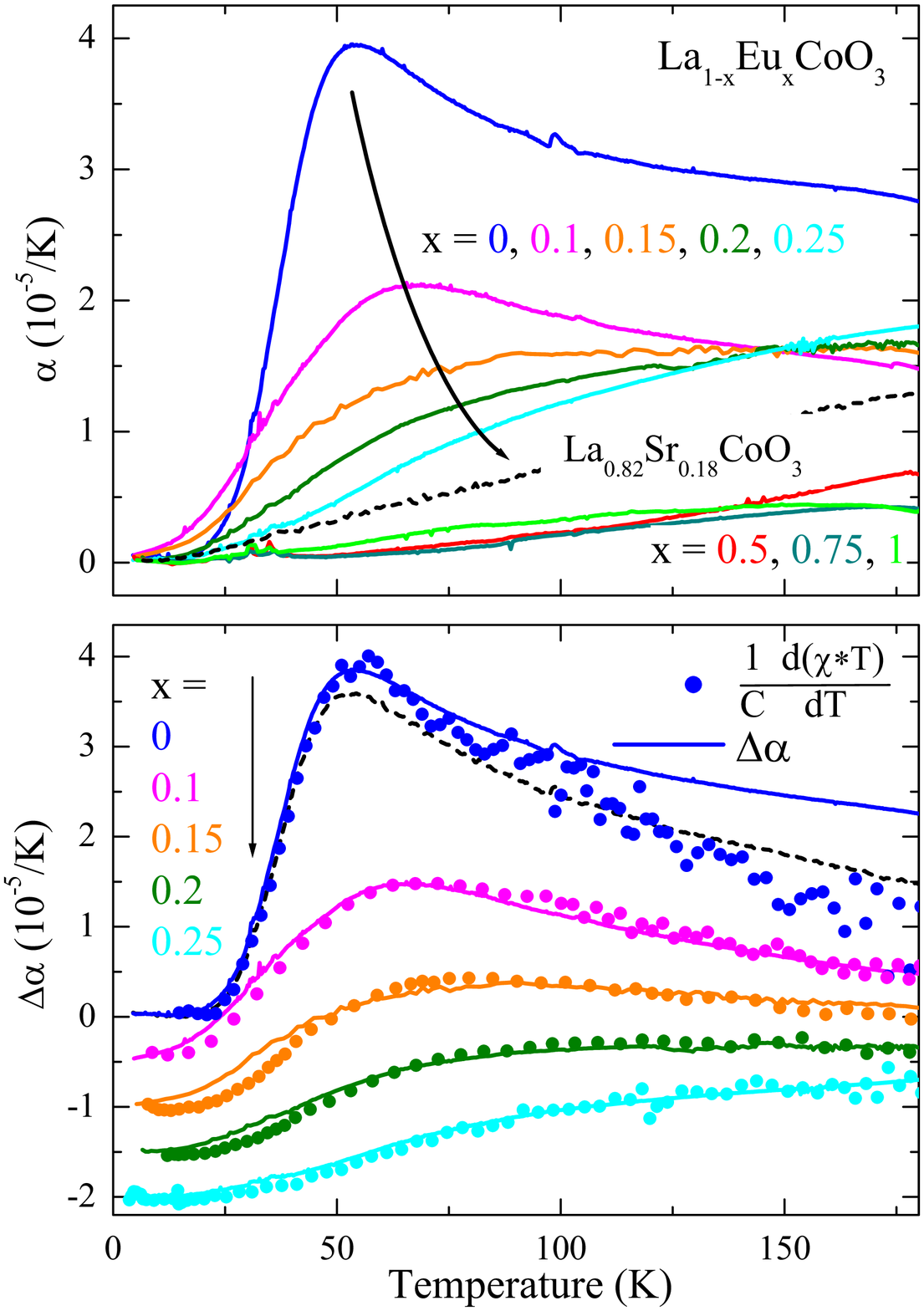}
\caption{(Upper panel) Thermal expansion $\alpha$ of
La$_{1-x}$Eu$_x$CoO$_3$ (solid lines) and
La$_{0.82}$Sr$_{0.18}$CoO$_3$ (dashed line). (Lower panel) Anomalous
thermal expansion $\Delta\alpha = \alpha-\alpha_{\rm bg}$ of \leco\
for $0\le x\le0.25$ (solid lines), where the averaged $\alpha$ of
\leco\ with $0.5\le x\le 1$ has been used as $\alpha_{\rm bg}$. The
dashed line shows $\Delta \alpha$ of \lco\ which was analysed in
Ref.\,\onlinecite{zobel02a} using $\alpha$ of
La$_{0.82}$Sr$_{0.18}$CoO$_3$ as $\alpha_{\rm bg}$. The symbols
($\bullet$) are obtained from the Co$^{3+}$ susceptibility data (see
lower panel of Fig.\,\ref{fig:chichico}) via the scaling relation
$C\cdot\Delta\alpha(T)=\partial(\chi_{\rm Co}\cdot T)\,/\,\partial
T$ with scaling factors $C$ between 190 and 210\,emuK/mole (see
Table\,\ref{tab:scaling}).  For clarity, the curves for different Eu
contents have been shifted by $-0.5\cdot 10^{-5}$/K with respect to
each other. \label{fig:thermex1}}
\end{figure}

The raw \tad\ data of \leco\ shown in the upper panel of
Fig.\,\ref{fig:thermex1} reveal an anomalous \tad\ of the samples
with $x\le 0.25$, which systematically flattens and shifts towards
higher temperatures with increasing $x$. In contrast, the
$\alpha(T)$ curves of the samples with $x\ge 0.5$ are essentially
identical and do not show any anomalous behavior up to 180\,K. It is
therefore reasonable to use the average of these curves as a
background $\alpha_{\rm bg}$ representing the normal lattice
expansion, which is present irrespective of the \sst . The anomalous
thermal expansion for $x<0.5$ is then obtained by $\Delta \alpha =
\alpha -\alpha_{\rm bg}$.\cite{alpha_back} As shown in the lower
panel of Fig.\,\ref{fig:thermex1} we find that for all samples with
Eu contents $0\le x\le 0.25$ the scaling relation between \tad\ and
magnetic \ch\ [Eq.\,(\ref{scala})] is well fulfilled. The scaling
factors for the different samples lie in the range from 190 to 210
emu K/mole and agree well with the value expected for a population of
the IS state without orbital degeneracy and deviate strongly from
other possible scenarios (see Table\,\ref{tab:scaling}). Thus, we
conclude that, just as in \lco , the \sst\ in \leco\ for $x\le 0.25$
arises from a thermal population of the IS state without orbital
degeneracy. With increasing Eu content the spin gap is enhanced and
the \sst\ is therefore shifted towards higher temperatures.

\begin{table} [t] %[H] add [H] placement to break table across pages
 \caption{Experimental results of the scaling factors $C$ in the scaling
relation $C\cdot\Delta\alpha(T)=\partial(\chi_{\rm Co}\cdot
T)\,/\,\partial T$ between anomalous thermal expansion
$\Delta\alpha$ and magnetic \ch\ $\chi$ for different Eu contents
$x$ (see Fig.\,\ref{fig:thermex1}). As shown in Ref. 15 the expected
values are C $simeq$ 190 emu K/mole for a thermal population of the
IS state without orbital degeneracy and $simeq$ 290 emu K/mole for
the IS state with orbital degeneracy. For a population of the HS
state with and without orbital degeneracy the expected values amount
to 1000 and 690 emu K/mole, respectively \cite{remark_C}.
 \label{tab:scaling}}
 \vskip1mm
 \parbox{7cm}{
 \begin{ruledtabular}
 \begin{tabular}{lccccc}
 x & 0 & 0.1 & 0.15 & 0.2 & 0.25 \\
 \hline
 \vspace{3pt}
 $C$\,(emuK/mole)  & 195 & 205 & 210 & 210 & 190
 \end{tabular}
 \end{ruledtabular}}
 \end{table}

\section{Discussion}

%\subsection{Metal-Insulator Transition}

\begin{figure}
\includegraphics[width=8cm,clip]{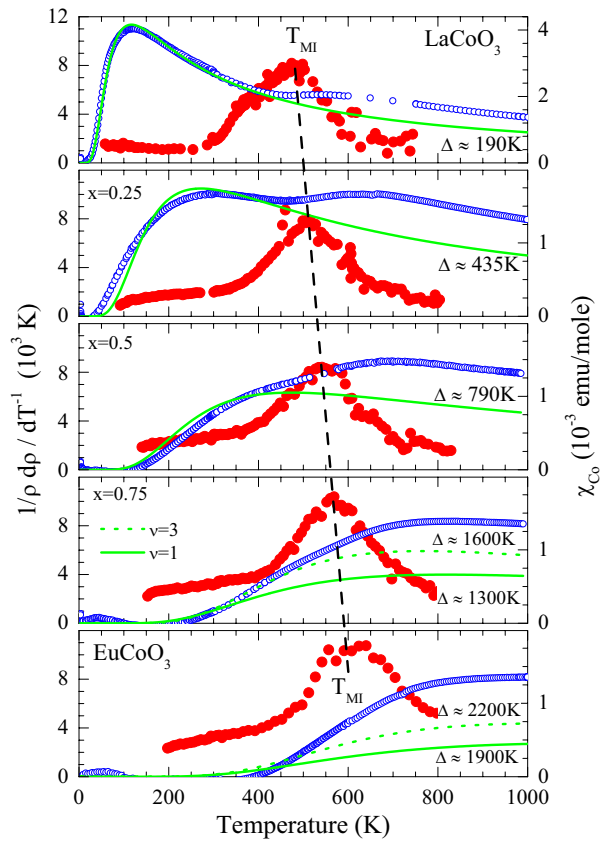}
\caption{Curie susceptibility of the Co$^{3+}$ ions (open circles,
right $y$ axis) in comparison with the temperature dependence of
the activation energy $\Delta_{\rm act} \simeq
\frac{\text{d}\ln\rho}{\text{d}(T^{-1})}$ (filled circles, left
$y$ axis). The peak of $\Delta_{\rm act}$ signals the \mi , which
slightly shifts towards higher temperatures as marked by the
dashed vertical line. The solid (dashed) lines are fits of
$\chi_{\rm Co}(T)$ sufficiently below \tm\ within a LS/IS scenario
with $\nu=1$ ($\nu=3$) (see Eq.\,\ref{twolevchi} and text).
\label{fig:chirho1}}
\end{figure}

In Fig.\,\ref{fig:chirho1} we compare the Curie susceptibility of
the Co$^{3+}$ ions obtained from the raw data after subtracting
$\chi_{\rm bg}$ [see Eq.\,(\ref{eq:background}) and
Table~\ref{tab:FitcDef}] and the resistivity which is presented in
the form $\frac{\text{d}\ln\rho}{\text{d}(T^{-1})}$. If this latter
quantity is constant, $\rho$ follows an activated behavior $\rho
\propto \exp\left(\Delta_{\rm act} / T\right)$ with a constant
activation energy $\Delta_{\rm
act}=\frac{\text{d}\ln\rho}{\text{d}(T^{-1})}$. At low temperatures
activated behavior is indeed observed. The broad peak of
$\frac{\text{d}\ln\rho}{\text{d}(T^{-1})}$ at higher temperatures
arises from the strong decrease of $\rho$ at the \mi . We find an
almost linear increase of the peak position, i.\,e.\ the \mi\
temperature \tm , with increasing $x$ from \tm ~$\simeq 480$\,K in
\lco\ to \tm ~$\simeq 600$\,K in \eco . For \lco\ $\Delta_{\rm act}
\simeq 1200$\,K is constant below room temperature.  With increasing
$x$ we find that $\Delta_{\rm act}$ deviates more and more from
being constant for $T<T_{\rm MI}$. This implies that for the samples
with larger resistivities the charge transport deviates from simple
activated behavior. Possibly it changes towards variable range
hopping, but the temperature range of our $\rho(T)$ curves is too
restricted to analyze this in more detail. If we ignore this
deviation and consider the room temperature values of $\Delta_{\rm
act}$ as a measure of an effective activation energy, we find a
strong increase of $\Delta_{\rm act}$ from 1200\,K for \lco\ to about
3400\,K for \eco . Obviously, the relative increase of the
activation energy is much more pronounced than that of the
transition temperature (see Table\,\ref{tab:gap}) as has also been
observed in RCoO$_3$.\cite{yamaguchi96b} In optical conductivity
data a strong similarity between the temperature-induced \mi\ of
\lco\ and the doping-induced \mi\ in \lsco\ has been
found.\cite{tokura98a} From this similarity and the anomalously
small \tm\ compared to $\Delta_{\rm act}$ it has been concluded that
the temperature-induced \mi\ of \lco\ should be viewed as a Mott
transition of a strongly correlated electron
system.\cite{yamaguchi96b,tokura98a}

Inspecting the susceptibility data in Fig.\,\ref{fig:chirho1} we
observe that the \mi\ of \leco\ is always accompanied by an increase
of $\chi$. This is obvious for the samples with $x<0.75$ where the
\sst\ and the \mi\ are well separated, whereas it is less obvious
for larger $x$, since the increase of $\chi$ due to the \sst\ and
due to \mi\ are superposed. In order to describe the \ch\ behavior
of \lco , various models have been considered. For the 100\,K \sst\
a thermal population of the HS or of the IS state has been proposed
and for both scenarios different orbital degeneracies $\nu$ have
been assumed for the excited state (HS: $\nu=1$ or 3; IS: $\nu=1$,
3, or
6).\cite{yamaguchi96a,saitoh97b,yamaguchi97a,asai98a,radaelli02a} In
order to also describe the change of $\chi$ around the \mi\ even
three-state scenarios (LS/IS/HS) with different orbital degeneracies
of the excited states have been
used.\cite{saitoh97b,asai98a,radaelli02a} In some cases a
temperature dependence of the energies of the different spin states
has also been considered, which can arise from the thermal expansion
and/or from collective interactions between the excited
states.\cite{asai98a,radaelli02a,kyomen03a,deltaofT}

As mentioned above the scaling between the \tad\ and the magnetic
\ch\ (see Fig.\,\ref{fig:thermex1} and Table\,\ref{tab:scaling})
gives clear evidence for a thermal population of the IS state with
$\nu=1$ and discards other two-state scenarios.\cite{zobel02a} This
unambiguous conclusion from the scaling analysis is only possible
for $x\le 0.25$ where the \sst\ occurs at low enough temperatures.
However, we will also use this scenario as a starting point for
larger $x$ and higher temperatures. In order to derive the energy
splitting $\Delta$ between the LS and IS state for the different Eu
concentrations we have fitted the respective $\chi_{\rm Co}(T)$ by
the \ch\ of a two-level system:
\begin{equation}
\chi(T) = \frac{N_{\rm A}g^2\mu_{\rm B}^2S(S+1)}{3k_{\rm B}T}\,
 \frac{\nu (2S+1)e^{-\Delta /T}}
 {1+\nu (2S+1)e^{-\Delta /T}}
 \;.
 \label{twolevchi}
\end{equation}
The first fraction describes the Curie \ch\ of the excited state and
the second one its thermal population with the Avogadro number
$N_{\rm A}$, the Bohr magneton $\mu_{\rm B}$, and the Boltzmann
constant $k_{\rm B}$, $\nu=1$ and $S=1$ denote the orbital
degeneracy and the spin of the IS state, which are kept fixed and the
$g$ factor is kept close to $g\simeq 2$.

\begin{table} [t] %[H] add [H] placement to break table across pages
 \caption{Activation energy $\Delta_{\rm act}$, \mi\ temperature
 \tm , energy splitting $\Delta$ between the LS and IS states,
 and $g$ factor of the IS state for different Eu contents
 (see Fig.\,\ref{fig:chirho1}).
 $\Delta_{\rm act}$ and \tm\ are deduced from the room temperature
 value and the maximum of $\frac{\text{d}\ln\rho}{\text{d}(T^{-1})}$, respectively.
 $\Delta$ and $g$ are obtained from fits of $\chi_{\rm Co}(T)$ in the
 temperature range below the \mi\ within a LS/IS scenario; for $x\le 0.5$ ($\ge 0.75$)
 an IS state without (with) orbital degeneracy has been considered. The last column
 contains the average energy gaps $\langle \Delta(T) \rangle$ for $T< 400$\,K obtained
 from the experimental data via Eq.\,\ref{eq:gap} with $g=2.4$
 (see Fig.\,\ref{fig:gap}.)  For
 $x\ge 0.75$ the values of $\Delta$ and $\langle \Delta(T) \rangle$ are only rough estimates,
 because $\chi_{\rm Co}(T)$ essentially vanishes
 in the respective low-temperature range and the agreement between the fits
 and the experimental data is rather weak.\cite{low_T}
 \label{tab:gap}}
 \vskip1mm
 \parbox{7cm}{
 \begin{ruledtabular}
 \begin{tabular}{lccccc}
 Eu content & $\Delta_{\rm act}$\,(K) & $T_{\rm MI}$\,(K) & $g$ &  $\Delta $\,(K)
 & $\langle \Delta(T) \rangle $\,(K) \\
 \hline
 \vspace{3pt}
 LaCoO$_3$ & 1200& 480 & 2.28 &  188 & 230\\
$ x=0.1  $ &  &        & 2.10 &  240 & 300\\
$ x=0.15 $ &  &        & 2.10 &  270 & 330\\
$ x=0.2  $ &  &        & 2.10 &  330 & 380\\
$ x=0.25 $ & 2000& 510 & 2.24 &  435 & 480\\
$ x=0.5  $ & 2800& 550 & 2.3  &  790 & 840\\
$ x=0.75 $ & 3200& 570 & 2.4  & 1600 & $> 1600$\\
 EuCoO$_3$ & 3400& 600 & 2.4  & 1900 & $> 2200$\\
 \end{tabular}
 \end{ruledtabular}}
 \end{table}

The fit parameters are summarized in Table~\ref{tab:gap} and the
corresponding fit curves are shown as solid lines in
Fig.\,\ref{fig:chirho1}. The fits  describe the experimental data
reasonably well up to about 450\,K for $x\le 0.5$. For larger $x$
the agreement between the fits and the experimental data becomes
worse, in particular for \eco . Because fits with $\nu =3$ yield
somewhat better results with larger gaps ($\Delta \simeq 1600$
and~2200\,K for $x=0.75$ and~1, respectively) one may speculate that
the orbital degeneracy is larger for $x\ge 0.75$. As mentioned above
the lifting of the orbital degeneracy in \lco\ arises from a JT
effect of the excited Co$^{3+}$ ions in the IS
state.\cite{zobel02a,maris03a} Since the \sst\ takes place at much
higher temperatures for $x\ge 0.75$, the JT effect could play a
minor role in these samples, since it is reduced by thermal
fluctuations. Moreover, the \sst\ sets in so close to the \mi\ that
the JT effect can hardly develop, since above \tm\ it will be
reduced anyway due to the enhanced charge-carrier mobility.
Independent of the choice of $\nu$ we find a drastic increase of the
energy gap $\Delta$ with increasing Eu content. We interpret this
increase as a consequence of the reduced unit-cell volume due to the
Eu substitution causing an enhanced crystal-field splitting which
stabilizes the LS relative to the IS state. In
Refs.\,\onlinecite{korotin96a,nekrasov03a,potze95a} it has been
argued that the IS state of Co$^{3+}$ can be stabilized by a
hybridization between the Co $3d$ and the O $2p$ states. Thus, the
increase of $\Delta$ may, in addition, be enhanced by the decreasing
Co $3d$ O $2p$ hybridization with increasing Eu concentration, which
is also reflected by the enhanced resistivity (see
Fig.\,\ref{fig:rho1}).

Before discussing the \ch\ data for higher temperatures let us
compare the influence of Eu substitution on the \sst\ and the \mi
. Obviously, the \sst\ is shifted much stronger to higher
temperatures than the \mi . This gives clear evidence that the
occurrence of the \mi\ in \leco\ is independent from the
population of the IS state of the Co$^{3+}$ ions. For example, at
$T_{\rm MI}\simeq 480$\,K \lco\ has almost a 3:1 distribution
between the IS and LS states according to their different
degeneracies, whereas in \eco\ the \sst\ just starts above 400\,K
so that the population of the IS state in \eco\ around $T_{\rm
MI}\simeq 600$\,K is as low as that in \lco\ around 50\,K. In
this respect the \sst\ and the \mi\ are decoupled from each
other. The observation that for all Eu concentrations $\chi_{\rm
Co}(T)$ seems to approach a common value for $T\rightarrow
1000$\,K indicates, however, that for temperatures well above the
\mi\ the same spin state (or combination of spin states) is
approached in \leco\ for all $x$.

As shown in Fig.\,\ref{fig:chirho1} $\chi_{\rm Co}(T)$ calculated
for the LS/IS scenario underestimates the \ch\ for $T\rightarrow
$\;\tm . Notably, well above \tm\ the deviations between the
extrapolations of the low-temperature fit curves and the
experimental data amount to about $5\cdot 10^{-4}$\,emu/mole for all
$x$ (if $\nu=3$ is used for $x\ge 0.75$). In \lco\ a well-defined
anomaly in the specific heat $C$ has been observed around \tm
,\cite{stolen97a} showing that the \mi\ represents a real phase
transition, in contrast to the \sst , which is a thermal population
of the excited IS state above the LS state causing only a broad
Schottky-type anomaly of $C$ (and $\alpha$) over a large temperature
interval. Consequently, the parameters (or even the model) for the
description of $\chi_{\rm Co}(T)$ above and below \tm\ can be very
different. In principle, there are various sources for an enhanced
\ch\ in the metallic phase. As mentioned above the charge-carrier
mobility above \tm\ should melt or at least reduce an orbital order
of the IS. Therefore, it is reasonable to use an orbital degeneracy
$\nu=3$ above \tm\ in Eq.\,(\ref{twolevchi}), which causes an
increase of $\chi_{\rm Co}(T)$.\cite{zobel02a,remark_C} However,
this effect is not sufficient to describe the enhanced \ch\ above
\tm\ as can be clearly seen from the dashed lines calculated with
$\nu=3$ in the lower panels of Fig.\,\ref{fig:chirho1}. From the
observation that charge-carrier doped \lmco\ ($M$\,=\,Ca, Sr, and
Ba) has a ferromagnetic, metallic ground state above a certain doping
level,\cite{kriener04a} one may also speculate that an additional
increase of $\chi_{\rm Co}(T)$ could arise from a ferromagnetic
coupling above \tm . However, \lco\ above \tm\ is not directly
comparable to \lmco , because the ferromagnetic coupling in \lmco\
arises from a double exchange between Co$^{3+}$ and Co$^{4+}$ ions,
whereas in \lco\ formally only Co$^{3+}$ ions are present.

A different source of an enhancement of $\chi_{\rm Co}(T)$ could
arise from a partial occupation of the HS state above \tm . Based on
this assumption describing the \ch\ of \lco\ within a LS/IS/HS
scenario has been attempted.\cite{saitoh97b,asai98a,radaelli02a}
However, even within this three-state model the \ch\ of \lco\ is only
reproduced if the energies of the IS and the HS state change with
temperature. Within this model a nearly temperature-independent
$\Delta_{\rm IS}\simeq 250$\,K and a strong decrease for
$\Delta_{\rm HS}$ from about $1400$ to about 200\,K occurring around
\tm\ has been reported for \lco .\cite{asai98a} Here, $\Delta_{\rm
IS}$ ($\Delta_{\rm HS}$) denote the energy between the LS and the IS
(HS) states. Due to the large value of $\Delta_{\rm HS}$ the HS
state is essentially not populated for $T<T_{\rm MI}$. This implies
that below \tm\ the \ch\ data of \lco\ are described within a LS/IS
model and above \tm\ within a LS/IS/HS model. Therefore, we have
also fitted $\chi_{\rm Co}(T)$ of our \leco\ samples separately
above and below \tm\ by the \ch\ of two- and three-state models.
However, in particular within the three-state model, there are so
many parameters that physically meaningful results can be obtained
only if a well justified model about the nature of the \mi\ and/or
the expected changes of the energies of the various spin states are
used.

In order to keep the number of parameters small we decided to
restrict ourselves to two-state models and will present a
description of $\chi_{\rm Co}(T)$ within a LS/IS scenario with
temperature-dependent energy gap as follows. An additional reason
why we do not use the three-state model is that the splitting
between different spin states is typically of the order of $1\,{\rm
eV}\simeq 12000$\,K.\cite{sugano70} The small splitting between the
LS and IS states ($\Delta \simeq 200$ - 2000\,K) observed in \leco\
already requires a fine tuning of parameters. Thus it appears quite
natural to assume that the HS state is located well above the other
two spin states. Once we have fixed the model in this way, it is
straightforward to derive the temperature dependence $\Delta(T)$ by
writing Eq.\,\ref{twolevchi} in the form
\begin{equation}
 \Delta(T) =  T\cdot \ln \left[
 \frac{N_{\rm A}g^2\mu_{\rm B}^2S(S+1)}{3k_{\rm B}T}\,
 \frac{\nu (2S+1)}{\chi(T)}
 -\nu (2S+1)\right]
 \,.
 \label{eq:gap}
\end{equation}

\begin{figure}[t]
\includegraphics[angle=270,width=8cm,clip]{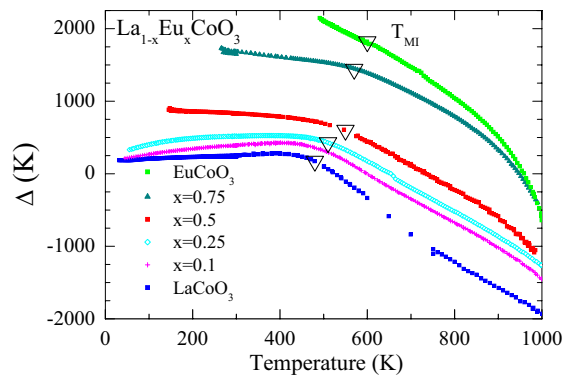}
\caption{Temperature dependence of the energy gaps $\Delta $
between the LS and the IS states obtained from Eq.\,\ref{eq:gap}
using $\nu=1$ for $x\le 0.5$ and $\nu=3$ for $x\ge 0.75$ and
$g=2.4$ for all $x$. (see text) \label{fig:gap}}
\end{figure}

Initially we tried to use the $g$ factors of Table~\ref{tab:gap}
obtained from the \ch\ fits well below \tm\ with temperature
independent energy gaps in order to derive the energy dependence
$\Delta(T)$ for larger temperatures. However, for $g < 2.4$ the
argument of the logarithm in Eq.\,(\ref{eq:gap}) becomes negative
above about 500\,K. Therefore we have used $g=2.4$ for all samples
and obtained the temperature-dependent energy gaps shown in
Fig.\,\ref{fig:gap}.\cite{low_T} For $x\le 0.5$ we have used $\nu=1$
and find a moderate temperature dependence of $\Delta (T)$ in the
low-temperature range. The average values $\langle \Delta (T)\rangle
$ for $T<400$\,K are close to the values obtained from the fits with
temperature-independent $\Delta$ [see Table~\ref{tab:gap}]. This
further justifies the fits of $\chi_{\rm Co}(T)$ within the LS/IS
scenario in the low-temperature range. For $x\ge 0.75$ we use
$\nu=3$ (as in the fits of $\chi_{\rm Co}(T)$) and find some
tendency to saturation only for $x=0.75$, but not for \eco . Thus,
for $x\ge 0.75$ we can give only lower boundaries of $\Delta(T)$,
which are so large that the obtained low-temperature values of
$\chi_{\rm Co}(T)$ are smaller than our experimental uncertainty. So
far the $\Delta(T)$ curves for $x\ge 0.75$ do not contradict the
LS/IS scenario for the low-temperature range.

At higher temperatures the $\Delta (T)$ curves strongly decrease for
all samples. This decrease sets in when the \mi\ is approached and
is present up to 1000\,K. A possible interpretation of the strong
decrease of $\Delta$ around \tm\ is that due to the \mi\ the IS
state is lowered in energy and is finally located well below the LS
state. Within this interpretation the continuous decrease of
$\Delta(T)$ up to the highest temperatures is not expected. One
should, however, keep in mind that the application of
Eqs.\,(\ref{twolevchi}) and~(\ref{eq:gap}) is questionable in the
conducting state above \tm , because these equations are based on a
model with localized moments. Note that this problem is also present
for the three-state models.

One may suspect that a temperature-dependent energy gap could
resolve the problem that on the one hand ESR data of \lco\ propose a
thermal population of a spin-orbit coupled HS state with
$\tilde{j}=1$ and $g\simeq 3.5$,\cite{noguchi02a}  while on the
other hand the \ch\ data are not at all reproduced by a population
of such a state with a temperature-independent gap. Therefore we
also calculated $\Delta(T)$ using these values in
Eq.~(\ref{eq:gap}). For \lco\ this leads to a continuously
increasing energy gap from $\Delta(T\le 40\,{\rm K})\simeq 200$\,K
to $\Delta(1000\,{\rm K})\simeq 1400$\,K interrupted by a plateau of
$\Delta$ from $ 500\,{\rm K} < T < 600$\,K, i.\,e.\ the temperature
dependence of $\Delta$ roughly resembles that of $1/\chi$. With
increasing Eu content $\Delta (1000\,{\rm K})$ remains essentially
unchanged, because $\chi (1000\,{\rm K})$ is almost constant, but the
temperature dependence of $\Delta $ continuously decreases until
$\Delta (T)$ is roughly constant for $x=0.75$ and finally decreases
with increasing temperature for $x=1$. Because there is no reason
why the temperature dependence of $\Delta$ should change so
drastically as a function of Eu content, we do not consider this
complex $\Delta(T,x)$ scenario as physically meaningful.

In conclusion, the analysis of the \ch\ data of \leco\ reveals that
a good description is possible within a LS/IS scenario below the \mi
. We find a drastic increase of the energy of the IS state with
increasing Eu content and that for all Eu concentrations the \mi\ is
accompanied by an increase of the \ch\ of about $5\cdot
10^{-4}$\,emu/mole. This increase either indicates a strong decrease
of the energy of the IS state around \tm , or that a model of
localized spins is no longer applicable above \tm . It remains to be
clarified whether $\chi_{\rm Co}(T)$ above \tm\ has to be described
by local moments or not.

\section{Summary}

We have studied the influence of chemical pressure on the spin-state
transition and on the \mi\ of \lco\ by partially substituting the
La$^{3+}$ by the smaller Eu$^{3+}$ ions. On the one hand the Eu
substitution drastically influences the activation energy for charge
transport in the insulating phase, which increases from about
1200\,K in \lco\ to about 3200\,K in \eco . The transition
temperature of the \mi , on the other hand, shows only a moderate
change from about 480\,K in \lco\ to about 600\,K in \eco . Contrary
to the \mi , the \sst\ is shifted very strongly towards higher
temperatures with increasing Eu content as is observed independently
in \tad\ as well as in magnetic \ch\ measurements of \leco . Both
quantities fulfill a simple scaling relation for $x\le 0.25$ and
$T\le 180$\,K. The experimentally obtained scaling factors agree
very well with the value expected for a thermal population of the IS
state without orbital degeneracy. Within this LS/IS scenario the
experimental \ch\ data below about 450\,K are satisfactorily
reproduced for $x\le 0.5$, whereas for larger $x$ an orbital
degeneracy of $\nu=3$ yields a better description. The lack of
orbital degeneracy for $x\le 0.5$ is an indirect evidence of orbital
order, or a JT effect, between the thermally excited Co$^{3+}$ ions
in the IS state. For $x \ge 0.75$ this effect seems to play a minor
role. This may arise from two effects which both act against an
orbital/JT ordering: the \sst\ occurs (i) at higher absolute
temperatures and (ii) closer to the \mi . With increasing Eu content
we find a drastic increase of the energy splitting between the LS
and the IS state from about 190\,K in \lco\ to about 2200\,K in \eco
. As a consequence the population of the IS close to \tm\ is very
different for both compounds: $\approx 70$\,\% in \lco\ and $\approx
25$\,\% in \eco . This difference clearly shows that the occurrence
of the \mi\ is not related to the population of the IS state. So far
the \mi\ and the \sst\ are decoupled from each other. The fact that
for all Eu concentrations the \ch\ increases by about $5\cdot
10^{-4}$\,emu/mole around the \mi\ and approaches a
doping-independent value for $T\rightarrow 1000$\,K indicates that
nevertheless a common spin state, or combination of spin states, is
approached well above \tm . The nature of this spin state remains,
however, unclear at present.

\begin{acknowledgments}
We acknowledge fruitful discussions with M.\,Braden,
J.\,B.\,Goodenough, M.\,Gr\"{u}ninger, E.\,M\"{u}ller-Hartmann,
D.\,Khomskii, and L.\,H.\,Tjeng. This work was supported by the
Deutsche Forschungsgemeinschaft through SFB~608.
\end{acknowledgments}

%\bibliographystyle{apsrev}
%\bibliographystyle{y:/bst/efk}
%\bibliography{y:/preload,y:/LaCoO3,y:/Schichtkobaltate,y:/ACoO3,y:/TMO,y:/Lehrbuch,y:/LaSrCoO3,../sonstiges}

\begin{thebibliography}{41}
\bibitem[{OOv()}]{OOvsJT}
\bibinfo{howpublished}{Although orbital order and JT effect
differ in their driving forces, an experimental distinction is
extremely difficult or even impossible because both effects are
strongly coupled to each other. In any material, orbital order
will be accompanied by a lattice distortion which lowers the
symmetry of the crystal field and enhances the splitting between
the orbitals. Vice versa, a collective JT effect gives rise to a
spatial ordering of the occupation of the orbitals.}
\bibitem[{IIv()}]{IIvsMI}
\bibinfo{howpublished}{Since for the Eu-doped samples the
resistivities above \tm\ amount up to 100\,m$\Omega$cm one could
label this transition an insulator-insulator transition. However,
there is no qualitative change of the high-temperature
transitions as a function of $x$. Thus we still label this a \mi\
as it is usually done in the case of \lco . }
\bibitem[{rem({\natexlab{a}})}]{remarkCovV}
\bibinfo{howpublished}{The low-temperature value of the van Vleck
  susceptibility of the Co$^{3+}$ ions ($\chi_{\rm vV}^{\rm Co}$) is two orders
  of magnitude smaller than that of the Eu$^{3+}$ ions. Thus, $\chi_{\rm
  vV}^{\rm Co}$ is expected to be constant in the examined temperature range,
  since due its smaller absolute value the temperature dependence of $\chi_{\rm
  vV}^{\rm Co}$ should also set in at much larger temperatures than that of
  $\chi_{\rm vV}^{\rm Eu}$, which starts to decrease above about 100\,K.}
\bibitem[{fit()}]{fitrange}
\bibinfo{howpublished}{We mention that this result does practically not change,
  when we use different temperature intervals for the fit, e.g. 4\,K $<T <$
  200\,K or 4\,K $<T <$ 400\,K.}
\bibitem[{Eug()}]{Eugap}
\bibinfo{howpublished}{From the fits in the doping range $0.5\le x\le 1$ we
  have found $457\,{\rm K} \le \Delta_{\rm Eu}\le 473$\,K and fixed this
  parameter to $\Delta_{\rm Eu} = 460$\,K for smaller $x$.}
\bibitem[{rem({\natexlab{b}})}]{remark_C}
\bibinfo{howpublished}{Because of the lower value of $\chi_0$ of LaCoO$_3$ the
  present fit of the susceptibility data yields a $g$ factor of 2.28 compared
  to $g=2.13$ used in Ref.~\onlinecite{zobel02a}. As one consequence the
  expected scaling factors $C$, which are proportional to $g^2$, increase by
  about 15\,\% compared to the values given in Table\,I of
  Ref.~\onlinecite{zobel02a}, e.g. from $\simeq 167$\,emuK/mole to $\simeq
  192$\,emuK/mole for the LS/IS scenario without orbital degeneracy. Another
  consequence is that increasing the orbital degeneracy from $\nu=1$ to $\nu=3$
  is not sufficent anymore to explain the enhanced \ch\ above \tm .}
\bibitem[{alp()}]{alpha_back}
\bibinfo{howpublished}{In Ref.\,\onlinecite{zobel02a} we have used the \tad\ of
  the charge-carrier doped La$_{0.82}$Sr$_{0.18}$CoO$_3$ as $\alpha_{\rm bg}$
  (dashed line in Fig.\,\ref{fig:thermex1}), which is somewhat larger than the
  $\alpha_{\rm bg}$ used here. Since $\alpha$ of \lco\ is much larger than both
  choices of $\alpha_{\rm bg}$, this uncertainty does hardly change $\Delta
  \alpha $ of \lco\ (upper solid and dashed line in the lower panel of
  Fig.\,\ref{fig:thermex1}). Thus, the conclusions about the \sst\ of \lco\
  discussed in Ref.\,\onlinecite{zobel02a} remain unchanged.}
\bibitem[{low()}]{low_T}
\bibinfo{howpublished}{Note that the Co$^{3+}$ \ch\ is expected to vanish
  exponentially for $T\ll \Delta$ within any model discussed so far.
  Experimentally, $\chi_{\rm Co}(T)$ is obtained from the raw data by
  subtracting $\chi_{\rm bg}$ (see Eq.~\ref{eq:background} and
  Fig.~\ref{fig:chirho1}). Therefore $\chi_{\rm Co}(T)$ scatters around zero
  below a certain temperature and we can derive $\Delta(T)$ only for
  temperatures where $\chi_{\rm Co}(T)$ is significantly larger than the
  scatter around zero.}
\end{thebibliography}

\end{document}